\documentclass[aps,twocolumn,groupedaddress,floats,amssymb]{revtex4}
\usepackage[dvips]{graphicx}
\usepackage{bm}

\begin{document}

\title{Weak First-Order Superfluid--Solid Quantum Phase Transitions}

\author{Anatoly Kuklov ${^1}$,
Nikolay Prokof'ev $^{2,3}$, and Boris Svistunov $^{2,3}$}
\affiliation{
${^1}$ Department of Engineering Science and Physics, The
College of Staten Island, City University of New York, Staten
Island, New York 10314  \\
${^2}$ Department of Physics, University of
Massachusetts, Amherst, MA 01003 \\
${^3}$ Russian Research Center ``Kurchatov Institute'',
123182 Moscow
}

\begin{abstract}
We study superfluid--solid zero-temperature transitions in
two-dimensional lattice boson/spin models by Worm-Algorithm Monte
Carlo simulations. We observe that such transitions are typically
first-order with the exception of special high-symmetry points
which require fine tuning in the Hamiltonian parameter space. We
present evidence that the superfluid--checkerboard solid  and
superfluid--valence-bond solid transitions at half-integer filling
factor are extremely weak first-order transitions and in small
systems they may be confused with continuous or high-symmetry
points.
\end{abstract}

\maketitle

Recently, there has been an increased interest in
superfluid--solid (SF-S) quantum phase transitions in lattice
boson/spin systems \cite{Hebert,SS,Demler,Sachdev}. [By `solid' we
mean an insulating state featuring the broken translation
symmetry---like checkerboard solid/antiferromagnet (CB) or valence
bond solid (VBS), as opposed to the Mott insulator which
preserves the translation symmetry.] On one hand, this interest
is stimulated by experimental perspectives of studying such
transitions in optical lattices \cite{Demler}, on the other hand,
the SF--VBS transition in a $(2+1)$-dimensional system has been
argued to be the simplest example of qualitatively new type of
quantum criticality, that does not fit the standard
Landau-Ginzburg-Wilson paradigm \cite{Sachdev}. Intriguing data on
the SF-S transitions were obtained by direct Monte Carlo
simulations of  quantum systems \cite{Hebert,SS}. It was observed
\cite{Hebert} that the SF-CB transition in the hard-core bosonic
model with nearest- and next-nearest-neighbor interaction remains
remarkably insensitive to the explicit Heisenberg symmetry
breaking. The transition is numerically indistinguishable from
that of the Heisenberg spin-1/2 model: a degenerate
(hysteresisless) transition. In simulations of the 2D quantum XY
model with ring exchange \cite{SS}, the SF-VBS transition was
interpreted as the second-order one, which suggested its 
understanding in terms of the deconfined quantum critical 
point \cite{Sachdev}.

In this Letter, we perform a careful finite-size analysis of the
SF-CB and SF-VBS transitions by simulating a generalized
(2+1)-dimensional J-current model \cite{Wallin94}, which is a
discrete-imaginary-time analog of the quantum boson/spin lattice
system. The simplicity and flexibility of our model in combination
with the efficient Worm Algorithm allow us to arrive at a
definitive conclusion that in  SF-CB and SF-VBS cases we are
dealing with anomalously weak first-order phase transitions.
Moreover, the two transitions are remarkably similar to each
other. In both cases, small enough critical systems mimic the
behavior of a highly symmetric model with broken symmetry. The
SF-CB case corresponds to the O(3)-symmetry of the Heisenberg
model, while in the SF-VBS case we clearly see a quasi-O(4)
behavior that manifests itself as a coexistence at the critical
point of the superfluid response and both ($x$- and $y$-) VBS
orders, in arbitrary proportions.

We confine ourselves  to the case when the statistics of
(2+1)-dimensional particle trajectories in imaginary time
(worldline configurations) is positive definite. What groundstates
can emerge under this condition? The state with chaotic
(unstructured) typical worldline configuration is SF. Indeed, the
absence of structure implies fluctuations of the worldline winding
numbers, $W$, and thus a finite superfluid response which is given
by the mean square of $W$ \cite{Ceperley}, see Eq.~(\ref{rho})
below. [One can hardly extend this argument to cases when the
configuration weight is not positive definite, since the notion of
a typical configuration becomes vague.] In SF, the U(1) symmetry
is broken (at least in a topological sense). The only way to
restore this symmetry is to suppress fluctuations of $W$ which
seems to be impossible without structuring worldline
configurations in such a way that for each imaginary-time moment
the position of each worldline in the corresponding real-space
plane can be unambiguously associated at the microscopic level
with one of the lattice sites/bonds, and vice versa. If the total
number of the worldlines  is not equal to a multiple of the number
of sites/bonds, the structured worldline configuration immediately
implies a broken translation symmetry. This consideration leads to
the conjecture that for models with  the positive definite
statistics of worldline configurations and non-integer filling
factor the generic groundstate should feature an order, either SF
or solid, or both (supersolid). Groundstates with none symmetry
being broken (``quantum disorder") may then occur only as critical
points separating the ordered states.

In the path-integral representation, we see no qualitative
difference between the site- and bond-based solids since both are
characterized by the worldline structuring, in the above-mentioned
sense. In either case, zero-point fluctuations necessarily include
{\it permutations} of two and more lines, and thus on large scales
such microscopic details as the most probable positioning---on
sites or on bonds---of the worldlines can hardly be relevant to
the universal properties of the SF--S transition. The only
property that seem to be crucial is the symmetry of the emerging
lattice.

In terms of algorithmic simplicity and numerical efficiency,
classical $(d+1)$-dimensional analogs of $d$-dimensional quantum
systems offer a significant advantage. This approach was
successfully used previously in the studies of disordered
\cite{Wallin94,AS,disorder} and two-component systems
\cite{two-component}. In addition, classical models offer more
freedom in ``designing" effective models with complex phase
diagrams. The so-called J-current model of Ref.~\cite{Wallin94} is
obtained by considering trajectories in discrete imaginary time.
Let ${\bf n} = ({\bf x}, \tau ) $ denote points of the
$(d+1)$-dimensional space-time lattice, and  integer currents
$J_{\nu }({\bf n}) $ with $\nu = \hat{x}, \hat{y}, \hat{\tau }$
specify how many particles are going from site ${\bf n}$ in
direction $\nu$. In this language, currents in the time-direction
represent occupation numbers, and currents in the space directions
represent hopping events.
 The continuity of trajectories requires that
 $\sum_\nu [J_{\nu }({\bf n})-J_{\nu }({\bf n}-\nu )]\equiv 0 $.

\begin{figure}[tbp]
\includegraphics[bb=0 0 520 800, angle=-90, width=1.\columnwidth]{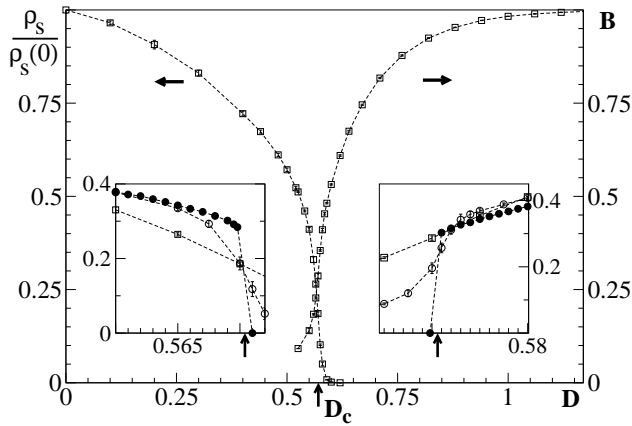}
\caption{ SF stiffness $\rho_s$ and VBS order parameter B
dependence on the coupling strength D for $L=16$ (squares), $L=32$
(open circles on inset) and $L=64$ (filled circles on inset)
systems. A closer look at the transition point is provided in the
insets. Error bars are shown for all points (in some cases these
are smaller than the symbol size).} \label{fig2}
\end{figure}

The simplest J-current model at half-integer filling factor is obtained by
writing the potential energy term in the particle-hole
symmetric form
\begin{equation}
S_{J}=2J \sum_{{\bf n},\nu \ne \hat{\tau}}
\left(J_{\hat{\tau}}({\bf n}) -{1\over 2} \right)
\left(J_{\hat{\tau}}({\bf n}+\nu ) -{1\over 2} \right)\;,
\label{AFM}
\end{equation}
and restricting currents along $\hat{\tau}$-bonds to take on just
two values, $0$ and $1$. The kinetic energy term is simply
\begin{equation}
S_{K}= K \sum_{{\bf n}, \nu \ne \hat{\tau }} |J_{\nu }({\bf n})|
\; , \label{Kin}
\end{equation}
with the restriction that allowed values for spatial currents are
$1$, $0$, and $-1$ values. To exclude somewhat pathological cases
with two hopping events happening at the same space-time point, we
further require that $ \sum_{\nu \ne \hat{\tau }} |J_{\nu }({\bf
n})|+|J_{\nu }({\bf n}-\nu )| \le 1 $. Finally, we introduce
interactions between the spatial currents on n.n. bonds which
favor VBS
\begin{equation}
S_{D}=-D \! \sum_{{\bf n}, \nu \ne \hat{\tau }} \!  |J_{\nu }({\bf n})|
\sum_{\mu \ne \nu }\!  \left( |J_{\nu }({\bf n}+\mu)|+ |J_{\nu }({\bf n}-\mu )|
 \right) .
\label{VB}
\end{equation}
Equal-time coupling ($\mu \ne \hat{\tau }$) favors simultaneous
hopping events of particles on the same plaquette  and is
reminiscent of the ring exchange term in quantum models \cite{SS}.
Phonon mediated exchange is another known mechanism of
dimerization in spin Pierles systems \cite{SSH}, and in
Eq.~(\ref{VB}) it is represented by coupling between bonds
connecting the same sites and shifted in the time direction. With
all three terms combined, $S=S_{J}+S_{K}+S_{D}$, the resulting
model has SF, CB, and VBS states in its phase diagram. 

First we study the SF-VBS transition along the $J=0$, $K=0.4$
line. Superfluid stiffness is determined by the statistics of
winding number fluctuations \cite{Ceperley}
\begin{equation}
\rho_s=\langle W^2\rangle /2L \;, \label{rho}
\end{equation}
and the VBS order parameter is characterized by the staggered
distribution of spatial currents along $\hat{x}$ and $\hat{y}$
directions
\begin{equation}
B_{\nu } = L^{-(d+1)} \sum_{\bf n} \, |J_{\nu }({\bf n})|\,
e^{i{\bf nq}} \;, \label{dimer}
\end{equation}
where $\nu = \hat{x}$ and $\nu =\hat{y}$ for ${\bf q}=(\pi,0,0)$
and ${\bf q}=(0,\pi ,0 )$, respectively. We find it convenient to
introduce a single VBS order parameter with positive definite
estimator which takes on finite value $\sim {\cal O}(1)$ in the
VBS phase, $B = |B_{\hat{x}}|+ |B_{\hat{y}}| $. For completeness,
we also define here the CB order parameter as
\begin{equation}
M_{\tau } ({\bf q}=(\pi, \pi ,0)) = L^{-(d+1)} \sum_{\bf n}
J_{\hat{\tau } }({\bf n}) e^{i{\bf nq}} \;. \label{M}
\end{equation}
and $M= |M_{\tau } | $.

In Fig.~\ref{fig2} we show rescaled data for the SF stiffness
$\rho_s/\rho_s(D=0)$ and VBS order parameter. The main plot for
$L=16$ demonstrates strong suppression of $\rho_s$ and $B$ near
the critical point [$D_c \approx 0.5705(2)$] which is typical for
continuous phase transitions. Similar behavior was reported
previously for the ring-exchange model in Ref.~\cite{SS}. However,
in the insets we clearly see that finite-size scaling is
incompatible with the second-order transition scenario---the
curves $\rho_s(D)$ for different sizes $L$ intersect each other
without any further rescaling indicating that large systems are
more ordered in the vicinity of the critical point. Similar
behavior is observed for the curves $B(D)$. The most obvious
scenario is then a first-order transition where the intersection
of finite-size curves at the critical point is allowed.
Apparently, the transition is {\it weakly} first-order because (i)
both order parameters are strongly suppressed at $D_c$, and (ii)
simulations for $L^3=32^3$ system do not show any hysteresis,
though the autocorrelation time is very long at $D_c$.

\begin{figure*}[tbp]
\begin{widetext}
\includegraphics[bb=50  95 600  815, angle=-90, width=0.66\columnwidth]{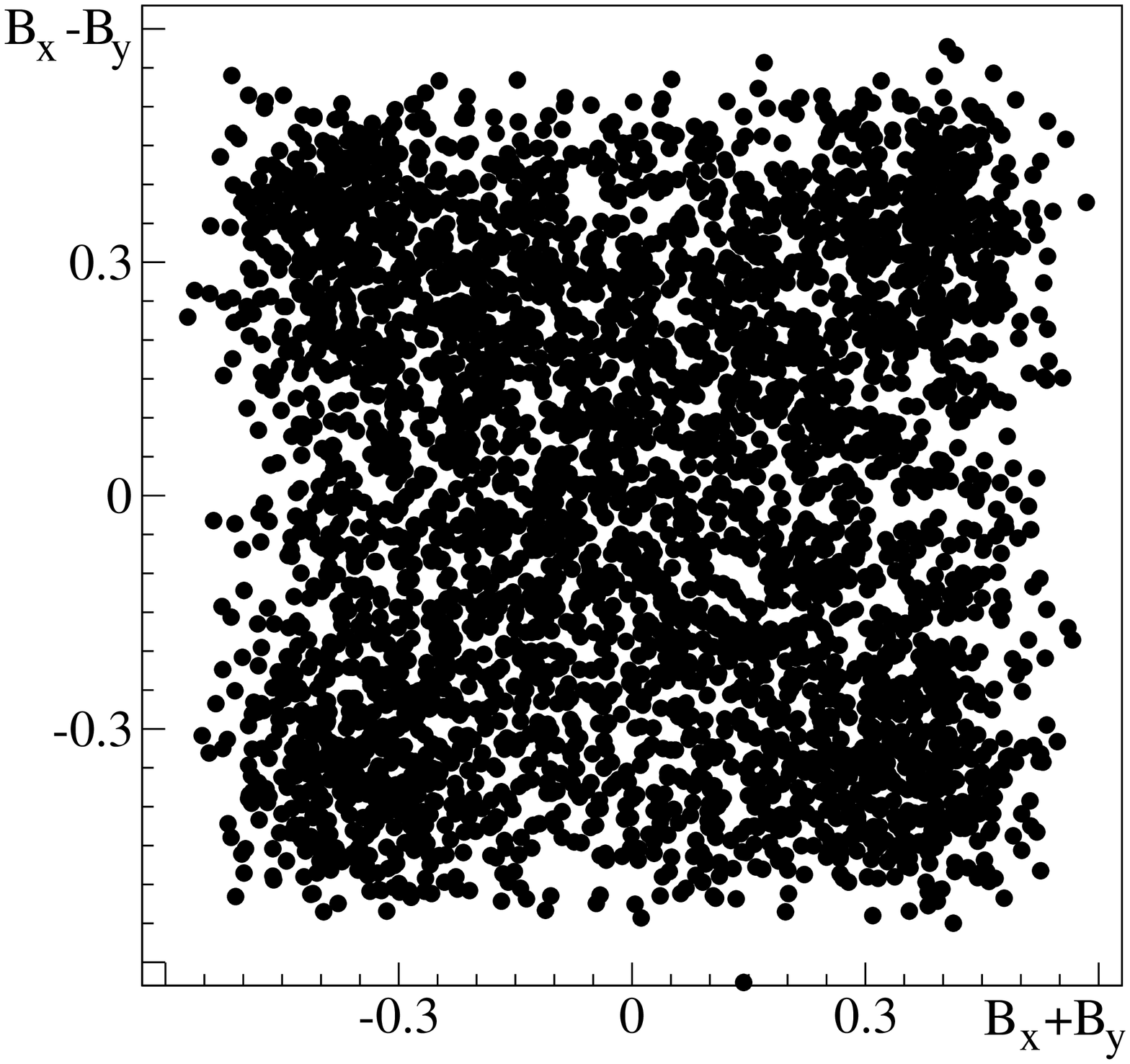}
\includegraphics[bb=40  100 600 840, angle=-90,width=0.66\columnwidth]{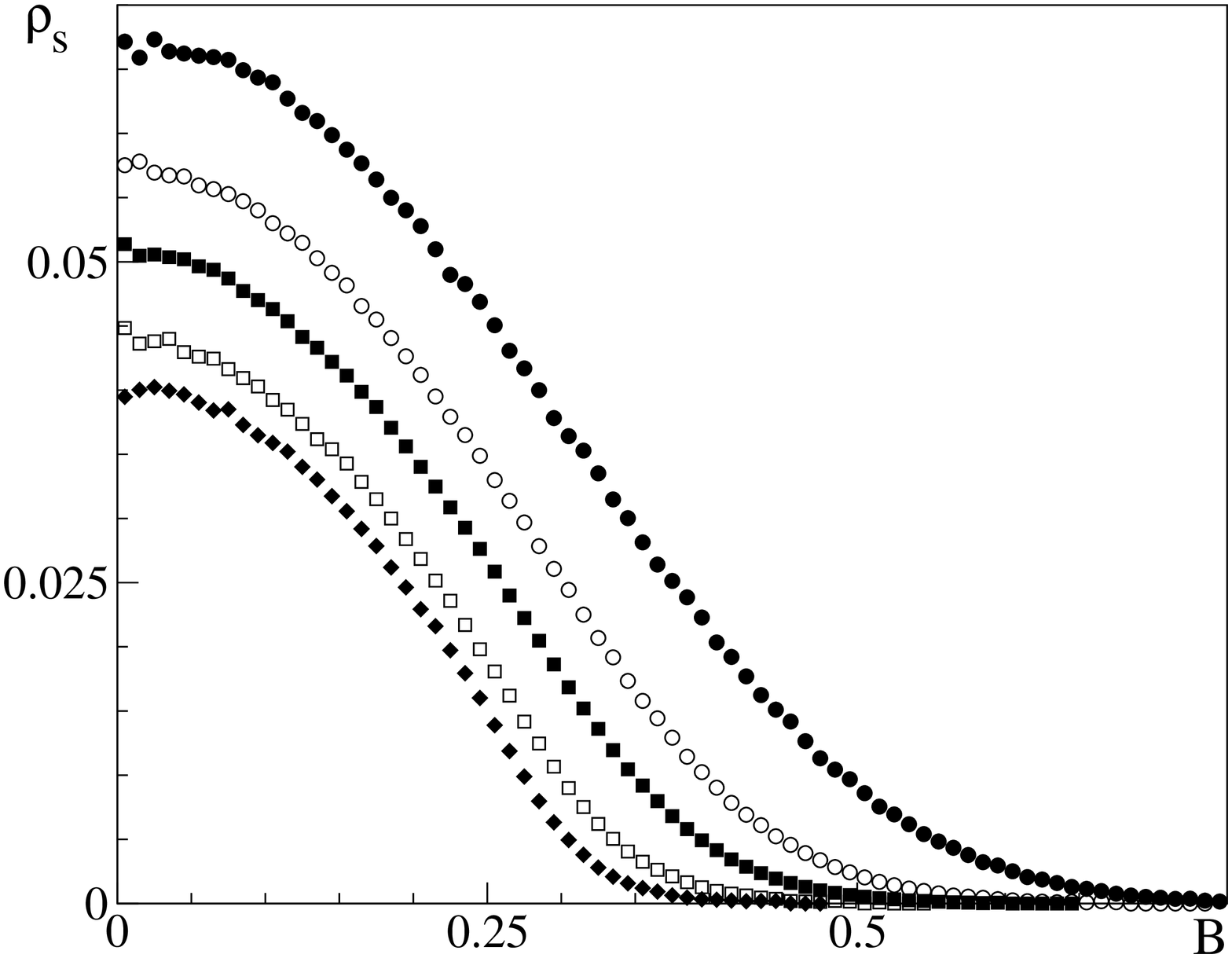}
\includegraphics[bb=40  20 600  760, angle=-90, width=0.66\columnwidth]{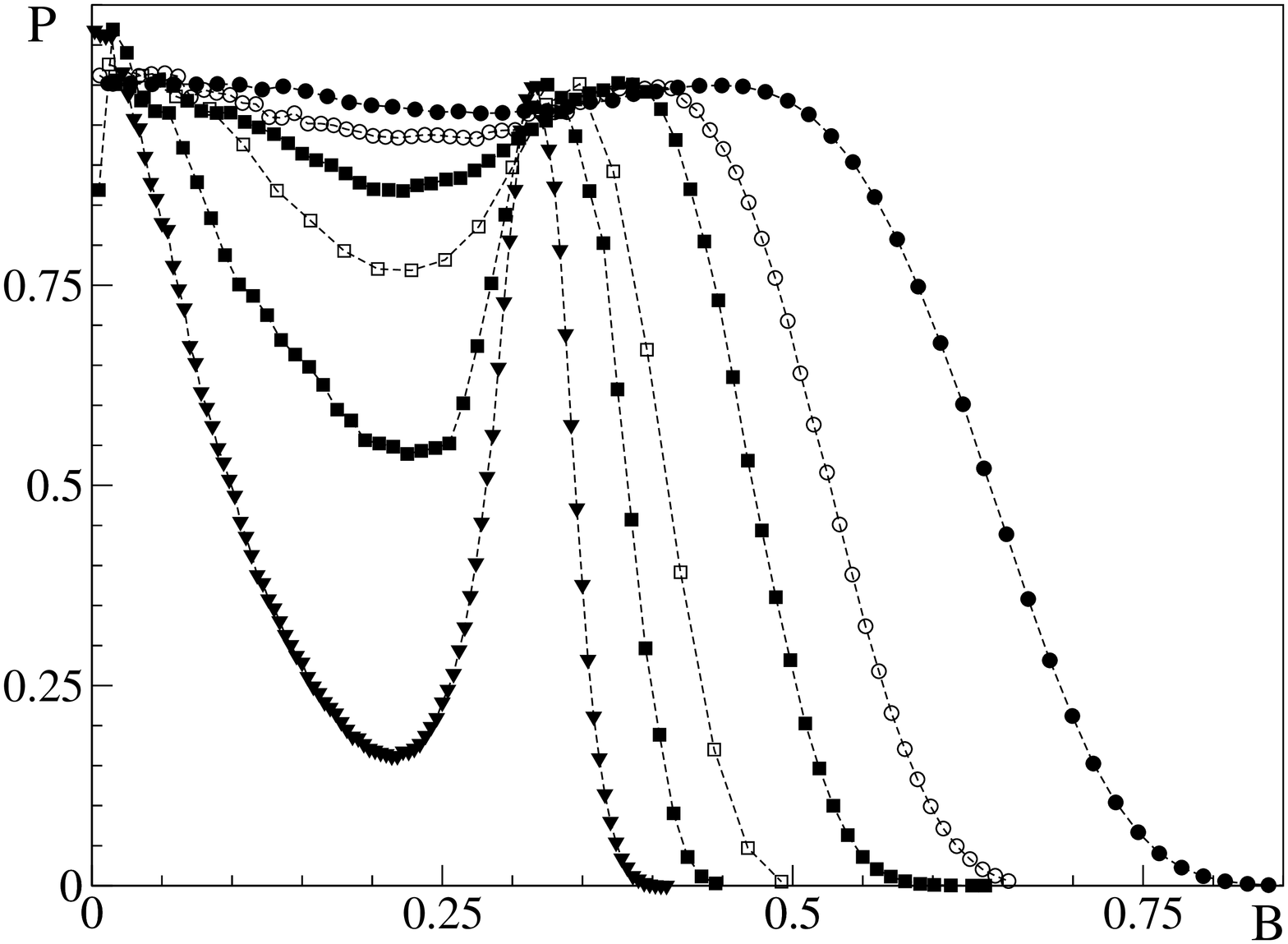}
\caption{Statistics of fluctuating SF and VBS order parameters.
Left panel: points of the ``black square'' are obtained for MC
configurations separated by equal long time intervals for $L=16$
and $D=0.5727$. Middle panel: SF stiffness is calculated as a
function of $B$ for various system sizes. From top to bottom
curves are plotted at corresponding finite-$L$ ``transition
points" (see the text), for $(L=8,~D=0.5742)$, $(L=12,~D=0.5738)$,
$(L=16,~D=0.5727)$, $(L=24,~D=0.5715)$, $(L=32,~D=0.5712)$,
$(L=48,~D=0.5705)$. Right panel: the coarse-grained distribution
of the average density of points in the $ B_{\hat{x}}, B_{\hat{y}}
$-square along the $B=|B_{\hat{x}}|+|B_{\hat{y}}|$=const lines for
the same set of system sizes and values of $D$ as in the middle
panel.} \label{fig3}
\end{widetext}
\end{figure*}
The other surprising fact is that the region where $\rho_s$ for
$L=16$ is below the corresponding curves for $L=32$ and $L=64$  is
rather extended, while in first-order transitions it is expected
to be macroscopically small. It appears as if the superfluid order
parameter $\Psi $ experiences anomalously large fluctuations in
small systems. To explain it we first speculate (and later prove
numerically) that in the vicinity of the critical point the system
is best described by the four-dimensional order parameter,
$\vec{S}$, and the O(4)-symmetry is broken at $D_c$ . Formally,
$B_{\hat{x}}$, $B_{\hat{y}}$ and two components of $\Psi $
represent observable (linearly independent) projections of
$\vec{S}$; correspondingly, in the $ ( B_{\hat{x}}, B_{\hat{y}},
\Psi )$-space the O(4)-sphere is seen as a four-dimensional
surface with the sphere topology. In this scenario, if the
O(4)-symmetry is exact then any value of $\vec{S}$ is equally
probable at $D_c$, i.e. solid orders along both spatial direction
and superfluidity may coexist.

An analogous well-known example of the O(3)-symmetric point is
provided by the SF-CB transition in the 2D quantum
XXZ-antiferromagnet with n.n. exchange interactions. In this case,
the XY order parameter $S_x+iS_y \equiv \Psi $ and the CB order
parameter $S_z \equiv M$ form a three-dimensional vector
$\vec{S}$. The critical point itself is described by the
SU(2)-symmetric Heisenberg Hamiltonian. Since the O(3)-symmetry is
spontaneously broken in the ground state of the Heisenberg model,
at the transition both $\rho_s $ and $M$  change discontinuously
in the thermodynamic limit, but this change occurs without energy
barriers and is preceded by anomalously large fluctuations and
long autocorrelation times in finite systems.

If the outlined picture is correct, then small perturbations which
explicitly break the O(n) symmetry at the {\it microscopic } level
should result in generic weak first-order transitions. Indeed, in
the spontaneously ordered state all renormalizations are finite.
Thus symmetry breaking perturbations, which couple to the {\it
long-range} order, result in the non-flat macroscopic energy/effective
action profile for the order parameter. Phase transitions are, then,
between energy minima separated by macroscopic barriers which may,
however, remain relatively weak in small systems. The J-current
model studied here has no microscopic symmetries mandating exact
O(4)- or O(3)-symmetry of the critical point. We conclude then
that SF-VBS and SF-CB transitions are expected to be first order,
and in the rest of the paper we present evidence that this is
indeed so.

\begin{figure}[tbp]
\includegraphics[bb=0 30 590 830, angle=-90, width=1.\columnwidth]{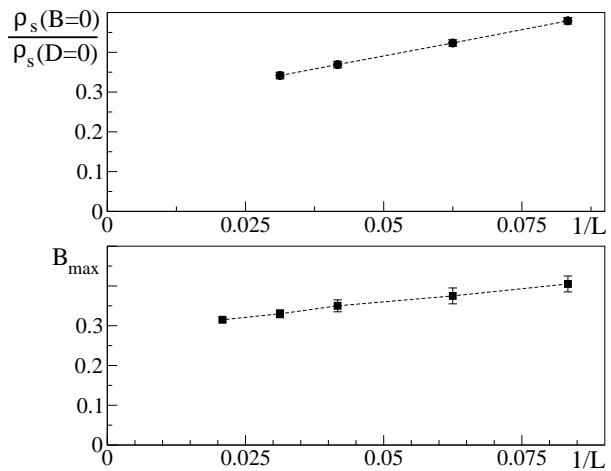}
\caption{System size dependence of $\rho_s(B=0)/\rho_s (D=0)$
(circles) and $B_{\rm max}$ (squares) from Fig.~\ref{fig2}. }
\label{fig4}
\end{figure}
\begin{figure}[tbp]
\includegraphics[bb=20 0 570 800, angle=-90, width=1.\columnwidth]{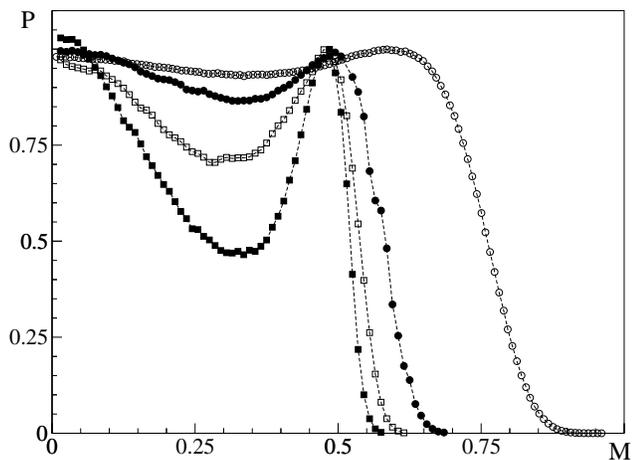}
\caption{The coarse-grained distribution of the CB order parameter
for various system sizes; from right to left: $(L=8,~J=0.450)$,
$(L=16,~J=0.4205)$, $(L=24,~J=0.4189)$, $(L=32,~J=0.4186)$. }
\label{fig5}
\end{figure}

First, we demonstrate that relatively ``small'' (hundreds of
particles!) systems  in the vicinity of the critical point behave
as if $ ( B_{\hat{x}}, B_{\hat{y}}, \Psi )$ fluctuations are
confined to some surface, not volume, and it is not possible to
have all three order parameters fluctuating to zero. In the left
panel of Fig.~\ref{fig3} we show statistical fluctuations of the
VBS order parameters for $L=16$ and $D=0.5727$. The density of
points in the $ B_{\hat{x}}, B_{\hat{y}} $ ``black square" is
nearly homogeneous, i.e. the system is equally likely to be found
with small or relatively large VBS order oriented at any angle
relative to $\hat{x},\hat{y}$-directions. Moreover, the boundary
shape suggests that $B_{\hat{x}}$ and $B_{\hat{y}}$ fluctuations
happen on the $B=f( |\Psi | )$ surface. The middle panel of
Fig.~\ref{fig3} shows that indeed the values of $B$ and $\rho_s$
are strongly correlated [$\rho_s (B)$ is defined as the average,
see Eq.~(\ref{rho}), over configurations with the VBS order
parameter $\in (B-\delta,B+\delta )$]. We see that, $\rho_s$ is
largest when $B \to 0$, i.e. despite large fluctuations of all
order parameters, they never vanish simultaneously.

Finally, in the right panel we plot the average density of points
in the $ B_{\hat{x}}, B_{\hat{y}} $-plane along the $B$=const cuts
for various system sizes. In each case, we adjusted the coupling
constant $D$ so that the distribution function $P(B)$ is
``maximally flat", or the two maxima are at equal heights.
Normalization was set to have the large-B maximum equal unity. The
density profiles are nearly flat for $L=6,12,16$ though with a
tiny minimum between the VBS and SF regions. It appears as if
physical order parameters belong to some four dimensional surface
and may diffuse over it without large effective action barriers.

The minimum gets more pronounced for $L=24,32$, and breaks the
distribution into well separated peaks for $L>48$, i.e. the energy
barrier finally gets large enough to localize the order parameter
in one of the phases. The two-peak structure is a smoking gun
evidence that the transition is ultimately weak first-order.
Further evidence is provided by finite-size scaling of the
critical-point distribution functions, i.e the dependence of
$\rho_s(B=0,L)$ (normalized to $\rho_s(D=0)$ as in
Fig.~\ref{fig2}) and the position of the $P(B)$ distribution
maximum, $B_{\rm max}(L)$. The data in Fig.~\ref{fig4} suggest
that both quantities saturate to finite values in the
thermodynamic limit.

The study of the SF-CB transition along the $K=0.7$, $D=0$ line
reveals a remarkable similarity with the SF-VBS point. At
$J_c=0.4184(2)$ the superfluid and solid orders exchange places
with  pronounced fluctuations of both order parameters in small
systems. These fluctuations are almost identical to what is
expected for the O(3)-symmetric Heisenberg point where the
distribution function for the staggered order parameter,
$P(M<M_{\rm max})=P(S_z)$ is a step function. However, in larger
systems a minimum in $P(M)$ is developed. The double-peak
structure of $P(M)$ for $L=32$ leaves no doubt that we are
actually dealing with the weak first-order transition. Apparently,
system sizes in the study of the SF-CB point in Ref.~\cite{Hebert}
were too small to see the first-order transition.

At certain conditions (which with J-current model can be easily
achieved by adding appropriate terms) the supersolid (SS) phase
may intervene between SF and S phases. In this case, the vicinity
to a (quasi-)O(3)/O(4) symmetric point with broken symmetry may
render the SF-SS and SS-S transitions also first order, while
normally one might expect them to be of the second order.

In conclusion, we note that weak first-order transitions discussed
here can hardly be an artifact of the J-current model since they
reveal themselves on large space-time scales.

\end{document}